# Photocatalytic Properties of Anisotropic β-PtX$_2$ (X= S, Se) and Janus β-PtSSe monolayers

Pooja Jamdagni[1*], Ashok Kumar[2], Sunita Srivastava[1], Ravindra Pandey[3*]

and K. Tankeshwar[1*]

[1] *Department of Physics and Astrophysics, Central University of Haryana, Mahendragarh, India, 123031*

[2] *Department of Physics, Central University of Punjab, Bathinda, India, 151401*

[3] *Department of Physics, Michigan Technological University, Houghton, MI, USA, 49931*

(August 27, 2022)

**Emails:** j.poojaa1228@gmail.com (Pooja Jamdagni); pandey@mtu.edu (Ravindra Pandey); drtankeshwar@gmail.com (K. Tankeshwar)

**Abstract**

The highly efficient photocatalytic water splitting to produce clean energy requires novel semiconductor materials to achieve high solar-to-hydrogen energy conversion efficiency. Herein, the photocatalytic properties of anisotropic β-PtX$_2$ (X=S, Se) and Janus β-PtSSe monolayers are investigated based on density functional theory. Small cleavage energy for β-PtS$_2$ (0.44 J/m$^2$) and β-PtSe$_2$ (0.40 J/m$^2$) endorses the possibility of their mechanical exfoliation from respective layered bulk material. The calculated results find β-PtX$_2$ monolayers to have an appropriate bandgap (~1.8-2.6 eV) enclosing the water redox potential, light absorption coefficients (~10$^4$ cm$^{-1}$), and excitons binding energy (~0.5-0.7 eV), which facilitates excellent visible-light driven photocatalytic performance. Remarkably, an inherent structural anisotropy leads to the anisotropic and high carrier mobility (up to ~5 x 10$^3$ cm$^2$V$^{-1}$S$^{-1}$) leading to fast transport of photogenerated carriers. Notably, the small required external potential to derive hydrogen evolution reaction and oxygen evolution reaction processes with an excellent solar-to-hydrogen energy conversion efficiency of β-PtSe$_2$ (~16%) and β-PtSSe (~18%) makes them promising candidates for solar water splitting applications.

# 1. INTRODUCTION

To achieve highly efficient photocatalytic water splitting together with significantly reducing the environmental and energy concern, great efforts have been made to explore the novel two-dimensional (2D) materials capable of harvesting sunlight with high solar-to-hydrogen energy conversion efficiency and the aptness of separation and migration of photoexcited carriers[1, 2]. A variety of novel 2D materials as photocatalysts have been proposed including $PdSe_2$[3], $PdSeO_3$[4], $SiP_2$[5], β-GeSe[6], β-SnSe[6], $BeN_2$[7], GaAs[8], $C_3S$[9], $C_3N_5$[10], g-CN[11], $Pd_3P_2S_8$[12], CuCl (1.61 V)[13], $AgBiP_2Se_6$[14], $Ga_2SSe$[15], $Ga_2S_3$[16], $HfX_3$ (X=S, Se)[17], Janus MXY (M=Mo, W; XY=S, Se, Te)[18-20], Janus CrXY(XY=S, Se, Te)[21, 22], Janus PdXY(XY=S, Se, Te)[23], Janus $M_2$XY(M=Ga, In; XY=S, Se, Te)[24, 25], Janus PdPSeX(X=O, S, Te)[26], Janus $Pd_4S_3Se_3$[27] and 2D honeycomb polymers[28, 29]. However, due to low energy conversion efficiency and high kinetic overpotential, only a few such 2D materials shows good photocatalytic properties for water splitting. Therefore, the researchers are still searching for novel water-splitting photocatalysts with improved performance.

In the family of 2D materials, noble metal dichalcogenides (NMDs) with the chemical formula $MX_2$ (M=Pd, Pt; X=S, Se, Te) have gained enormous attention due to their unique chemical and physical properties [30-32]. They exhibit highly anisotropic structures with different polymorphs[33-35], layer-dependent[36-38] and strain-tunable electronic structure[39], outstanding optoelectronic[40, 41], sensing[42], and photocatalytic properties[43, 44]. Furthermore, they can form more than one stable phase, such as T-phase with hexagonal lattice[45] and puckered pentagonal structure with an orthorhombic lattice[46, 47]. One phase can undergo a phase transition by external stimuli such as under high pressure[48, 49], uniaxial strain[50], and Li-adsorption[51].

Among the noble metal dichalcogenides, Pt-based layered structures such as $PtSe_2$ have emerged as a promising 2D material because of their experimental synthesis[52], novel physical, chemical, and photonic properties[53-55]. Pt dichalcogenides exhibit a 1T-type crystal structure with an in-plane covalent metal-chalcogen bond and interlayer van der Waals interaction between the triatomic (chalcogen-metal-chalcogen) layers[56-58]. Also, a new class of 2D materials known as "Janus structure" [59] shows distinctly different properties, including an intrinsic built-in electric field with many potential applications[19, 60, 61]. The successful fabrication of Janus PtSSe monolayer has also been achieved[62], which shows excellent electronic, optical, and photocatalytic properties[63-67]. In this study, we now consider a novel 2D phase of $PtX_2$ (X=S, Se), referred to as β-$PtX_2$, and their Janus structure performing a comprehensive investigation of their properties for photocatalytic water splitting.

## 2. COMPUTATIONAL DETAILS

Density functional theory (DFT) calculations and ab-initio molecular dynamics (AIMD) simulations as implemented in the VASP program package[68] were performed to predict the stability and electronic properties of these 2D materials. The exchange-correlation term was described in DFT calculations by generalized gradient approximation (GGA) within Perdew–Burke–Ernzerhof (PBE) formalism. The plane-wave basis set with cut-off energy of 500 eV and projector augmented wave (PAW) potentials were used in calculations. The conjugate gradient method with forces and energy tolerance values 0.001 eV/Å and $10^{-8}$ eV, respectively, was used to relax the crystal structures. The first Brillouin zone (BZ) integration was performed with a $20 \times 20 \times 1$ k-point grid of Monkhorst pack mesh. A vacuum of 20 Å perpendicular to the surface was used in the periodic supercell. For calculation of the bandgap, the hybrid HSE06 functional[69] was employed. Furthermore, the GW-method [70] was employed to calculate the electronic structure using the Wannier90 package[71]. The GW wavefunctions were further used to calculate absorption spectra by solving the Bethe-Salpeter equation (BSE). The convergence tests reveal that GW+BSE spectra can be accurately obtained using 300 eV cut-off energy, 16x16x16 k-points mesh, and 64 conduction bands (Figure S1, Supplementary Information).

## 3. RESULTS AND DISCUSSION

### 3.1 Crystal Structure of β-PtX$_2$ (X=S, Se) and β-PtSSe

The β-PtX$_2$ crystal exhibits trigonal symmetry with P3$_1$21 space group[72] with weakly interacting stacked layers along the z-direction (Figure S2). The atoms of each layer are positioned in a structure having 4- and 6-atoms rings similar to β-Te[73]. The lattice parameters of the bulk structure of β-PtX$_2$ (X=S, Se) are given in Table S1. The β-PtX$_2$ monolayers exhibit a (4-6)-ring structure with rectangular cells in the x-y plane. Monolayer contains a helical chain-like structure in the x-direction where each Pt atom in the chain makes a covalent bond with the four neighbouring X (X=S, Se) atoms. The chains of covalently bonded X-X (X=S or Se) atoms repeat in the y-direction (Figure 1). The lattice constants (a, b) of β-PtX$_2$ (X=S, Se) monolayers are slightly smaller than their bulk counterparts. The lattice parameters of the Janus β-PtSSe monolayer lie between its constituent's monolayer (Table 1).

**Table 1:** Lattice constants (*a, b*), cohesive energy ($E_c$), bond lengths (R), bond angles (Θ), and work function (Φ) of β-PtS$_2$, β-PtSe$_2$ and Janus β-PtSSe monolayers. The structural parameters of the previously reported 1T-phase are also given for comparison.

|  | β-phase | | | 1T-phase[61, 67] | | |
|---|---|---|---|---|---|---|
|  | β-PtS$_2$ | β-PtSe$_2$ | β-PtSSe | 1T-PtS$_2$ | 1T-PtSe$_2$ | 1T-PtSSe |
| *a* (Å) | 3.51 | 3.67 | 3.59 | a=b=3.57 | a=b=3.69 | a=b=3.63 |
| *b* (Å) | 4.32 | 4.59 | 4.47 |  |  |  |
| R (Å) | Pt-S=2.34 S-S=2.15 | Pt-Se=2.46 S-S=2.46 | Pt-S=2.35 Pt-Se=2.45 S-Se=2.32 | Pt-S=2.40 | Pt-Se=2.52 | Pt-S=2.43 Pt-Se=2.50 |
| Θ (degree) | Pt-S-S=102.8 Pt-S-Pt=97.2 S-Pt-S=82.8 | Pt-Se-Se=100.2 Pt-Se-Pt=96.5 Se-Pt-Se=83.5 | Pt-S-Se=101.5 Pt-S-Pt=99.5 Pt-Se-Pt=94.2 S-Pt-Se=83.1 | Pt-S-Pt=83.9 S-Pt-S=96.1 | Pt-Se-Pt=85.6 Se-Pt-Se=94.4 | Pt-S-Pt=93.3 S-Pt-S=84.8 Pt-Se-Pt=97.0 |
| Φ (eV) | 5.97 | 5.56 | 5.57, 5.94 | 6.07 | 5.36 | 5.31, 6.09 |
| $E_c$ (eV/atom) | 4.41 | 4.02 | 4.21 | 4.71 | 4.45 | 4.31 |

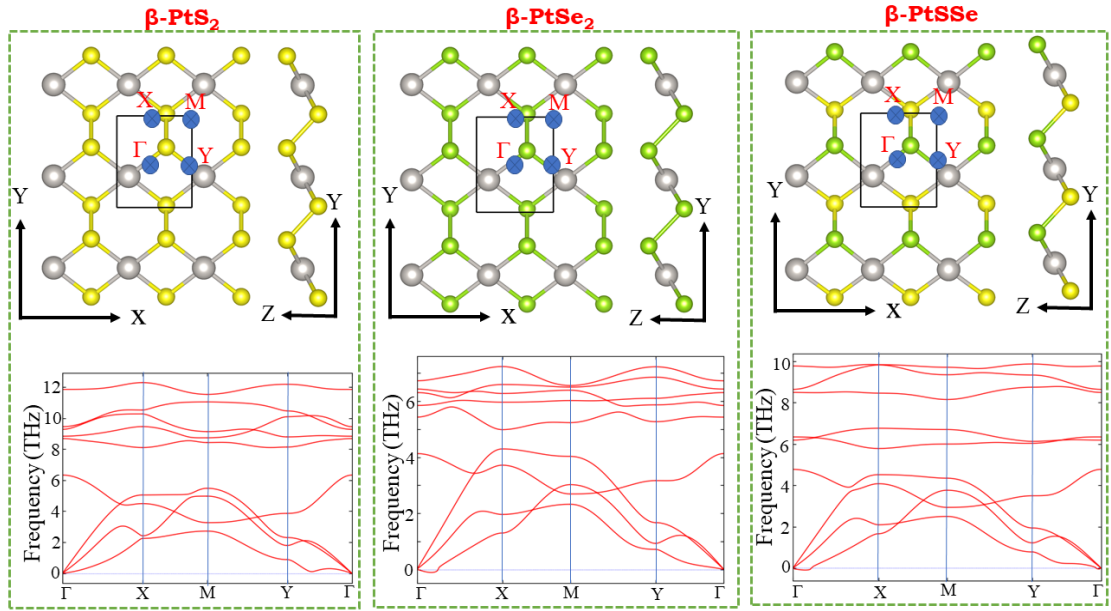

**Figure 1:** Top and side views of the monolayer configurations and phonon spectra of β-PtS$_2$, β-PtSe$_2$, and Janus β-PtSSe. The rectangular unit cell and high symmetry points in the corresponding Brillouin zone are also shown. Color code: Grey-Pt, Yellow-S, and Green-Se.

The symmetric electrostatic potential profile for β-PtS$_2$ and β-PtSe$_2$ monolayers and asymmetric potential as a function of the vertical distance for Janus PtSSe monolayer reveals its asymmetric out-of-plan crystal structure (Figure 2). The out-of-plane symmetry breaking in the Janus monolayer occurs due to the electronegativity difference between the chalcogen atoms that leads to an intrinsic built-in electric field of 0.87 eV/Å. Note that the slope of a line between the minimum of the planar average potential of outermost Se and S atoms can be used to calculate the built-in electric field[74]. The potential difference of 0.37 eV is calculated with respect to the two sides of the Janus PtSSe monolayer.

### 3.2 Stability Analysis

To assess the possibility of obtaining mechanically exfoliated monolayer β-PtX$_2$ from their bulk crystal, we calculate the cleavage energy as[75]:

$$E_{cl} = \frac{E_{iso} - E_{bulk}/n}{A} \quad (1)$$

where E$_{iso}$ is the energy of the unit cell of an isolated layer in a vacuum, E$_{bulk}$ is the energy of the bulk unit cell composed of n-layers, and A is the in-plane area of the bulk unit cell. E$_{cl}$ is calculated as 0.44 J/m$^2$ and 0.40 J/m$^2$ for β-PtS$_2$ and β-PtSe$_2$ monolayers, respectively. Remarkably, E$_{cl}$ of both monolayers is quite comparable with MoS$_2$ (0.42 J/m$^2$)[76] and PdSeO$_3$ (0.42 J/m$^2$)[4] and lower than Ca$_2$N (1.09 J/m$^2$)[77] monolayers, which have been realized experimentally via exfoliation method. Also, E$_{cl}$ of β-PtX$_2$ monolayers is lower than the theoretically investigated 2D monolayers such as GeS$_2$ (0.52 J/m$^2$)[78], GeP$_3$ (0.91 J/m$^2$)[79], α-BS (0.96 J/m$^2$)[80] and InP$_3$ (1.32 J/m$^2$)[81]. The cleavage energy calculations suggest the feasibility of experimental fabrication of β-PtX$_2$ (X=S, Se) monolayers.

Subsequently, we examine the energetic stability of β-PtX$_2$ and Janus PtSSe monolayers by calculating cohesive energy (E$_c$), defined as the difference between the total energy of the monolayer and the sum of the energies of the constituent isolated atoms. E$_c$ is calculated to be 4.41 eV/atom, 4.02 eV/atom, and 4.21 eV/atom for β-PtS$_2$, β-PtSe$_2$, and Janus β-PtSSe, respectively. The lower value of E$_c$ for selenides compared to sulphides is due to the larger atomic radius of Se than S, which produces a longer Pt-Se bond length and thus weaker bond strength as compared to the Pt-S bond. The E$_c$ of these monolayers is higher than the other 2D materials such as black phosphorene (3.79 eV/atom)[82], antimonene (2.75 eV/atom)[83], arsenene (3.19 eV/atom)[84], CuCl (3.04 eV/atom)[13] and SiP$_2$ (3.98 eV/atom)[5]. Our investigation demonstrates the possibility of the experimental fabrication of free-standing β-PtX$_2$ and Janus PtSSe monolayers.

The stability of β-PtX$_2$ (X=S, Se) and Janus PtSSe monolayers is further checked by calculating the formation energy ($E_f$), which indicates the energy difference during the formation of monolayers from their constituent elements. $E_f$ is defined as follows:

$$E_f(E_{PtX_2} \text{ or } E_{PtSSe}) = E_{PtX_2} (or\ E_{PtSSe}) - n_{Pt}E_{Pt} - n_X E_X (or - n_S E_S - n_{Se} E_{Se}) \qquad (2)$$

where $E_{PtX_2}$ and $E_{PtSSe}$ is the total energy of β-PtX$_2$ and Janus β-PtSSe unit cell, while $E_{Pt}$, $E_S$ and $E_{Se}$ is the energy of Pt, S and Se in their stable phases. Note that the space group of bulk Pt is $Fm\bar{3}m$ and bulk S and Se is $P3_121$. The formation energy is thus calculated as -0.78 eV/atom, -0.73 eV/atom and -0.72 eV/atom, respectively, for β-PtS$_2$, β-PtSe$_2$ and Janus β-PtSSe monolayers. The negative sign in the formation energy value indicates their energetic fabrication feasibility.

Next, the dynamic stability is analysed by calculating the phonon spectrum of β-PtX$_2$ and Janus β-PtSSe monolayers (Figure 1). The phonon spectrum is free from imaginary phonon modes revealing the dynamic stability of these monolayers. Also, AIMD simulations with a relatively larger 4 × 4 supercell at 500 K with a time step of 1 fs are performed to determine the thermal stability of these monolayers. After 10000 fs, the small total energy fluctuations with time and no structural distortion suggest that these monolayers are thermally stable (Figure S3) and show promises to the future synthesis of these monolayers.

### 3.3 Electronic and Optical Properties

After verifying the stability and feasibility of the experimental fabrication of β-PtX$_2$ and their Janus monolayers, we computed the electronic properties. β-PtS$_2$, β-PtSe$_2$ and Janus β-PtSSe monolayers are indirect gap semiconductor with bandgap of 1.56 eV (2.60 eV), 1.08 eV (1.77 eV) and 1.28 eV (2.09 eV) at GGA+PBE (HSE06) level of theories (Table 2). The valance band maximum (VBM) is located at Y-point (0.0, 0.5, 0.0), and the conduction band minimum (CBM) lies at Γ-point (0.0, 0.0, 0.0), as can be seen in Figure 2. of The electronic band structure computed at the GW-level of theory is shown in Figure S4. The indirect (direct) band gap at G$_0$W$_0$ level of theory is calculated to be 2.74 eV (2.80 eV), 2.02 eV (2.20 eV) and 2.28 eV (2.40 eV) for β-PtS$_2$, β-PtSe$_2$ and Janus β-PtSSe monolayers, respectively. The density of states (DOS) analysis reveals that the mix of Pt d-orbitals with the anionic p-orbitals contributes to the states near the Fermi level (Figure S5)

**Table 2:** Calculated band gaps ($E_g$) at different levels of models. The band gap values of the 1T-phase are also given for comparison.

|  | β-phase | | | 1T-phase[67] | | |
| --- | --- | --- | --- | --- | --- | --- |
|  | β-PtS$_2$ | β-PtSe$_2$ | β-PtSSe | 1T-PtS$_2$ | 1T-PtSe$_2$ | 1T-PtSSe |
| $E_g$ (eV), PBE | 1.56 | 1.08 | 1.28 | 1.81 | 1.34 | 1.56 |
| $E_g$ (eV), PBE+SOC | 1.50 | 0.94 | 1.15 | 1.77 | 1.15 | 1.41 |
| $E_g$ (eV), HSE06 | 2.60 | 1.77 | 2.09 | 2.78 | 1.88 | 2.28 |

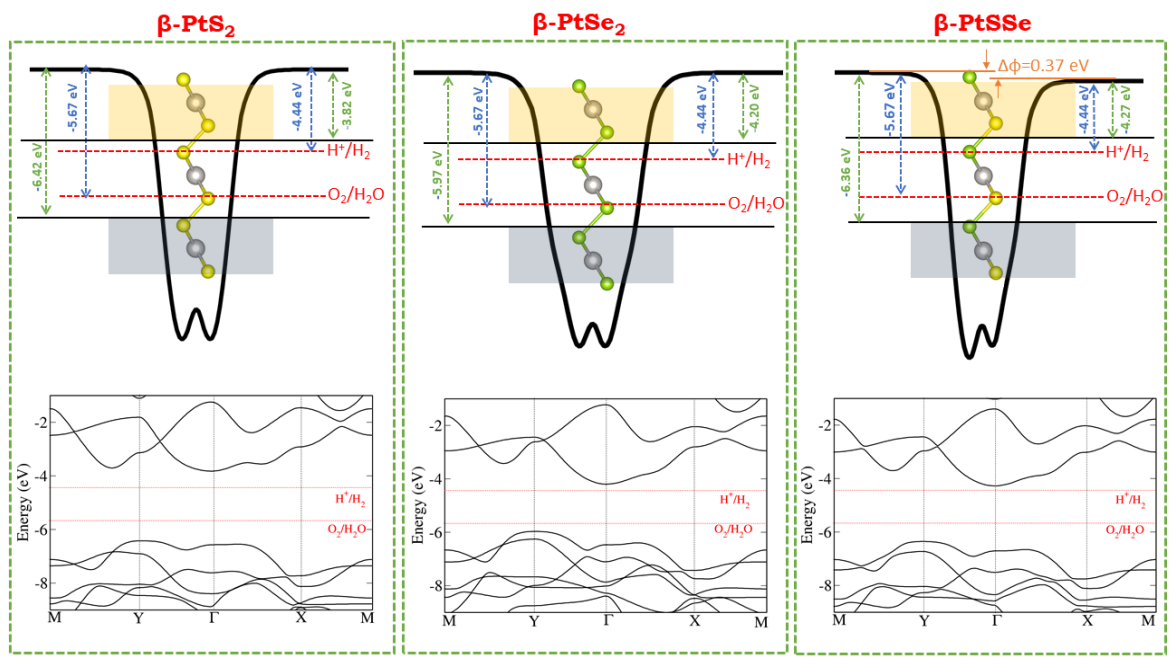

**Figure 2:** Calculated electrostatic potential profile and electronic band structure for β-PtS$_2$, β-PtSe$_2$, and Janus β-PtSSe at the HSE06 level of theory. Energy in the electronic band structure is aligned with respect to the vacuum level, at pH=0.

The effect of spin-orbit coupling (SOC) on the electronic band structure is investigated using GGA+SOC calculations. Our results reveal that the SOC effect on β-PtS$_2$ is negligible, whereas a small reduction in bandgap value ($\Delta E_g$ ~0.15 eV) is seen for β-PtSe$_2$ and Janus β-PtSSe monolayers (Figure S6).

It is generally accepted that 2D semiconductors with bandgap ranging from 1.23 eV to 3 eV are potential candidates for water-splitting photocatalysts. At the same time, the band edge position of VBM and CBM should be aligned with respect to the vacuum level relative to water oxidation potential ($E_{O_2/H_2O}$) and hydrogen reduction potential ($E_{H^+/H_2}$), respectively. As shown in Figure 2, the monolayers considered possess suitable band edge positions with respect to the standard redox potential, indicating that these monolayers satisfy the basic requirements for water splitting photocatalysts. Furthermore, β-PtS$_2$ is predicted to be a suitable photocatalyst for water splitting into acidic and neutral mediums. In contrast, monolayer β-PtSe$_2$ and Janus β-PtSSe possess band alignments for water splitting only in an acidic medium.

Also, an efficient photocatalyst must have excellent light-harvesting ability, particularly in the visible region. The optical response has been determined by calculating optical absorbance along in-plane (X and Y) direction as[85]: $A(\omega) = \frac{\omega}{c}\epsilon_2(\omega)L_z$ , where L$_z$ is the length of a supercell in z-direction, c is the speed of light, and ε$_2$ is the imaginary part of the dielectric function. The most prominent absorption bands lie within the visible region for all three monolayers (Figure 3), affirming the visible light-harvesting ability of these monolayers for solar water splitting. Also, the absorption coefficients (α) for all the monolayers in the visible region are calculated at ~10$^4$ cm$^{-1}$. The lowest energy transition associated with the excitons for β-PtS$_2$, β-PtSe$_2$, and β-PtSSe is calculated to be 2.07 eV, 1.67 eV, and 1.83 eV, respectively.

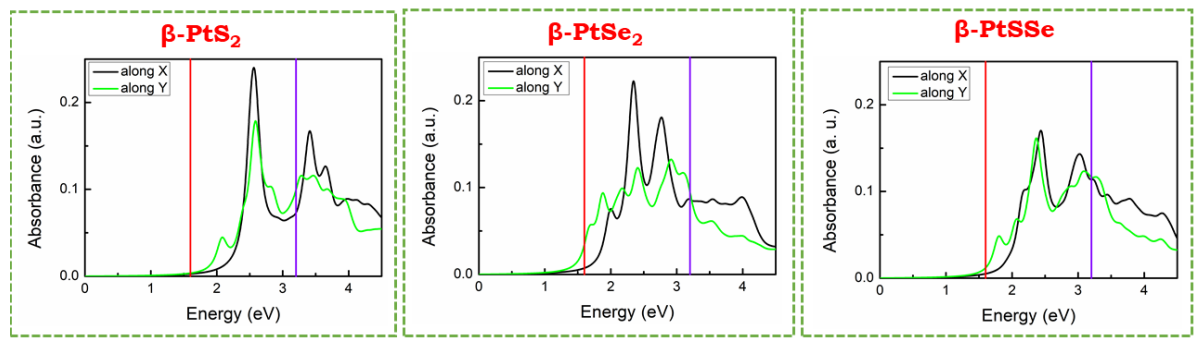

**Figure 3:** The optical absorbance of β-PtS$_2$, β-PtSe$_2$, and Janus β-PtSSe calculated at the GW+BSE level of theory.

Now we calculate the exciton binding energy to examine the photoexcitation-induced carrier separation. The exciton binding energy is calculated as E$_b$= E$_{QP}$-E$_{opt}$, where E$_{QP}$ and E$_{opt}$ are quasiparticle bandgap (E$_g^{GW}$, a minimal direct gap) and optical gap (i.e., the energy of the lowest transition in the optical spectrum), respectively. The excitons binding energy for β-PtS$_2$, β-PtSe$_2$, and β-PtSSe is calculated to be 0.73 eV, 0.53 eV, and 0.57 eV, respectively, which is less than or comparable with the excitons binding energy of Mo and W chalcogenides monolayers (~0.7-1.0 eV) [86]. Hence β-PtX$_2$ and their Janus monolayer have sufficient charge carriers for the reaction process of effective electron-hole pairs separation. Note that the excitons binding energy of selenide monolayers is lower than those of the sulphide monolayers due to enhanced dielectric screening provided by the heavier chalcogen. The value of static dielectric constant ($\varepsilon_s$) for β-PtS$_2$, β-PtSe$_2$, and β-PtSSe is calculated to be 2.73, 3.39, and 3.07, respectively.

### 3.4 Carrier Mobility

To explore the migration ability of photoexcited carriers, we now calculate the carrier mobilities by adopting well-accepted deformation potential theory and effective mass approximation[84] (see model details in supplementary information). High carrier mobility is desirable for efficient photocatalysis. Carrier mobility is determined by multiple parameters such as high elastic modulus, small deformation potential, and small effective mass generally gives high carrier mobility. The plots for elastic modulus and deformation potentials are given in Figures S7 and S8. The calculated relevant parameters and carrier mobility for β-PtS$_2$, β-PtSe$_2$, and β-PtSSe along X- and Y-direction are listed in Table 3.

**Table 3:** Elastic modulus (C$_{2D}$), deformation potential (E$_d$), effective mass (m*), and carrier mobility of β-PtS$_2$, β-PtSe$_2$, and Janus β-PtSSe monolayers along X-and Y-direction. Effective mass anisotropy (R$_m$*=m$_h$*/m$_e$*) and mobility anisotropy (R$_a$=μ$_{max}$/μ$_{min}$) are also given. Note that elastic modulus or in-plane stiffness of 1T-PtS$_2$, 1T-PtSe$_2$, and Janus 1T-PtSSe is previously calculated as 99, 88 and 93 in the units of Jm$^{-2}$ or N/m[61].

|  |  | β-PtS$_2$ | | β-PtSe$_2$ | | β-PtSSe | |
| --- | --- | --- | --- | --- | --- | --- | --- |
|  |  | X | Y | X | Y | X | Y |
| C$_{2D}$(Jm$^{-2}$) | | 152 | 72 | 129 | 44 | 127 | 56 |
| E$_d$ (eV) | hole | 0.73 | 0.85 | 3.15 | 1.62 | 2.36 | 1.93 |
| | electron | 0.70 | 3.59 | 1.0 | 3.35 | 1.1 | 3.16 |
| m*(m$_0$) | hole | 1.19 | | 2.94 | | 3.24 | |
| | electron | 0.88 | | 0.80 | | 0.77 | |
| μ (cm$^2$V$^{-1}$s$^{-1}$) | hole | 2873 | 997 | 21 | 28 | 31 | 20 |
| | electron | 5688 | 103 | 2877 | 87 | 2584 | 135 |
| R$_m$* | | 1.35 | | 3.68 | | 4.21 | |
| R$_a$ | hole | 2.9 | | 1.3 | | 1.6 | |
| | electron | 55 | | 33 | | 19 | |

According to our calculations, electrons mobility of all the monolayers is one order higher along X direction (~10$^3$ cm$^2$V$^{-1}$S$^{-1}$) as compared to Y-direction (~10$^2$ cm$^2$V$^{-1}$S$^{-1}$). The highly anisotropic electrons mobility is due to the highly anisotropic deformation potential and elastic modulus, e.g., deformation potential (elastic modulus) of β-PtS$_2$, β-PtSe$_2$ and β-PtSSe along X- and Y-direction is calculated as 0.70 eV and 3.59 eV (152 Jm$^{-2}$ and 72 Jm$^{-2}$), 1.0 eV and 3.35 eV (129 Jm$^{-2}$ and 44 Jm$^{-2}$), and 1.10 eV and 3.16 eV (129 Jm$^{-2}$ and 56 Jm$^{-2}$), respectively. The electrons mobility of these monolayers is large and much larger than that of MoS$_2$ (72.16 cm$^2$V$^{-1}$S$^{-1}$)[87] and WS$_2$ (130 cm$^2$V$^{-1}$S$^{-1}$)[88], demonstrating their fast carrier migration ability.

To get the extent of anisotropy in carrier mobilities, mobility anisotropy, defined as the ratio of maximum mobility to minimum mobility of a charge carrier, is calculated. The anisotropy in electron mobility is much higher than hole mobility. The anisotropy in the carrier mobility is due to the inherent structural anisotropy in β-PtX$_2$ and their Janus monolayer that leads to the directional-dependence elastic modulus and deformation potentials (Table 3). Also, the large difference in the effective masses (m$_e$* and m$_h$*) leads to different transfer mobility (μ=eτ/m*) and hence the different diffusion lengths [L$_p$=(μk$_b$τ/e)$^{1/2}$ = k$_b$τ$^2$/m*)$^{1/2}$] that lowers the carrier's recombination rate[89]. The relative L$_p$ can be quantitatively estimated by effective mass anisotropy (R$_m$*=m$_h$*/m$_e$*). The large calculated R$_m$* values (Table 3) suggest lower recombination rates of photogenerated carriers in these monolayers. Hence, the large mobility and effective mass anisotropy of carriers in β-PtX$_2$ and their Janus monolayer reduce the carrier recombination and improve the efficiency of these photocatalysts for water splitting.

**3.5 Gibbs Free Energy Profiles**

For efficient photocatalytic water splitting, the excellent electronic and optical properties must be supported with the feasible HER and OER mechanisms and processes (Gibbs free energy profiles) and high solar to hydrogen (STH) efficiency. Gibb's free energy calculations for the intermediates in HER and OER are performed in a 3x3 supercell. The details of computations can be found in the supplementary information. Note that Gibb's free energy has a difference of 100-150 meV on including vdW functional in the calculations. The photocatalytic water splitting is enhanced by the photogenerated carriers' induced external potentials. The calculated potential of the photogenerated electrons (holes) for hydrogen reduction (water oxidation) at pH = 0 for β-PtS$_2$, β-PtSe$_2$ and β-PtSSe is calculated to be 0.62 V (1.98 V), 0.24 V (1.53 V) and 0.17 V (2.30 V), respectively.

For the practical applications of photocatalysts, the stability of these monolayers in an aqueous solution needs to be investigated. Following the approach proposed in literature[90], the thermodynamical reduction potential (φ$^{re}$) and oxidation potential (φ$^{ox}$) are calculated, and the calculation details are given in supplementary information. The calculated value of $\phi^{re}$ relative to NHE for β-PtS$_2$, β-PtSe$_2$ and β-PtSSe monolayers are -1.37 V, -1.62 V and -1.49V, respectively, which are higher than the reduction potential for H$^+$/H$_2$. The calculated value of $\phi^{ox}$ is 2.41 V, 2.14 V and 2.27V, respectively, which are lower than the oxidation potential for $O_2/H_2O$. Therefore, photogenerated carriers have the tendency to react with water molecules rather than photocatalyst monolayers, which reveals that these monolayers are insoluble in an aqueous solution.

The HER mechanism on these monolayer surfaces follows the two-electron (2e) reaction pathway described recently for other 2D materials[19, 43, 44]. In the first step, the proton and an electron combine with a monolayer to form an H* species, which is not energetically favourable, having an HER barrier (without light irradiation, U=0) of 1.29 eV, 1.0 eV, and 1.13 eV for β-$PtS_2$, β-$PtSe_2$ and β-PtSSe, respectively. The H2 molecule is released in the second step by a favorable exothermic process (Figure 4). The position of band edges allows these monolayers to act as photocatalysts up to pH=7 for β-$PtS_2$, pH=3 for β-$PtSe_2$, and ph=2 for β-PtSSe. The photogenerated electron potential for hydrogen reduction reduces the HER barrier. However, the external potential of 0.67 V, 0.76 V, and 0.69 V at pH=0 is still required to derive the HER process spontaneously on the surface of β-$PtS_2$, β-$PtSe_2$, and β-PtSSe, respectively. These values are lower than the recently reported 2D photocatalysts materials for HER, such as $SiP_2$ (0.83 V)[5], PE-$AgBiP_2Se_6$ (1.06 V), and FE-$AgBiP_2Se_6$ (1.62 V)[14].

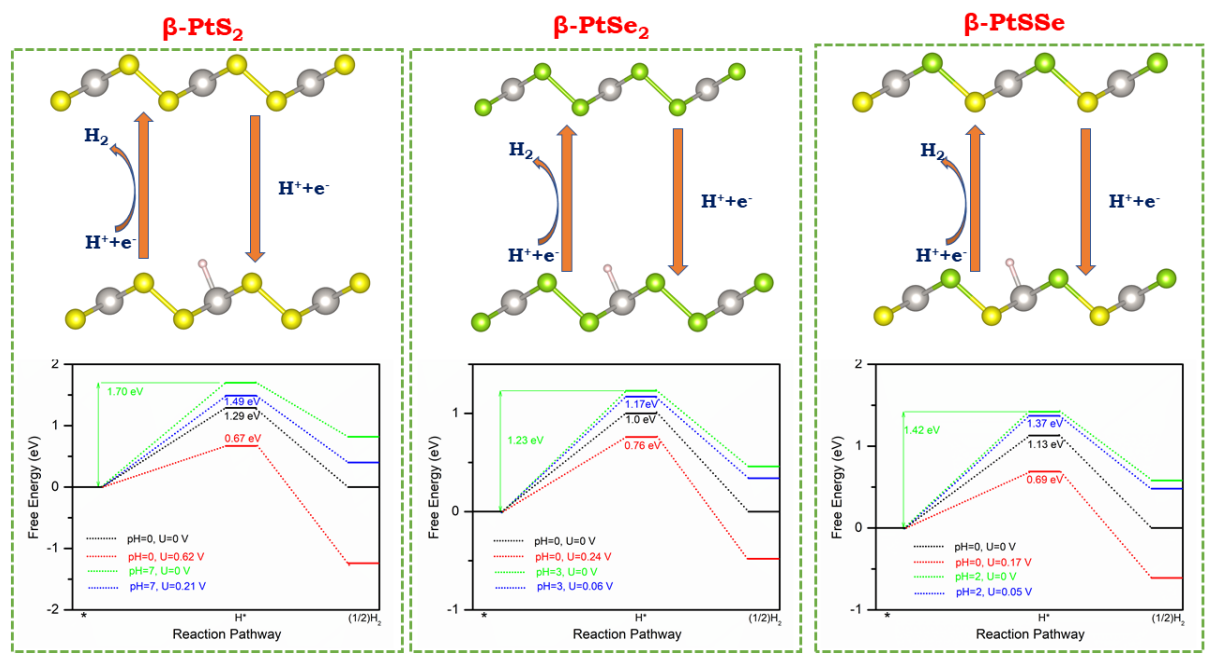

**Figure 4:** The photocatalytic pathways and free energy profiles of β-$PtS_2$, β-$PtSe_2$, and Janus β-PtSSe monolayers for HER at different pH and photogenerated electron potentials. Color code: Grey-Pt, Yellow-S, Green-Se, Light Pink-H.

The OER mechanism on the surface of these monolayers follows the four-electron (4e) reaction pathway (Figure 5). The rate-limiting potential for β-PtS$_2$, β-PtSe$_2$, and β-PtSSe at pH=0 is the formation of OOH* intermediate with the value of 8.46 eV, formation of OH* intermediate with the value of 2.39 eV and formation of OOH* intermediate with the value of 3.98 eV, respectively. These values (OER barrier) are reduced to 8.05 eV, 2.16 eV, and 3.69 eV by changing the solution conditions to pH=7 for β-PtS$_2$, pH=3 for β-PtSe$_2$, and pH=2 for β-PtSSe. The photogenerated hole potentials of 2.39 V for β-PtS$_2$, 1.71V for β-PtSe$_2$, and 2.42 for β-PtSSe further reduce the OER barrier. The external potential of 5.66 V, 0.45 V, and 1.27 V is required to derive the OER process spontaneously on the surface of β-PtS$_2$, β-PtSe$_2$, and β-PtSSe, respectively. Note that the recently reported 2D photocatalysts materials with required external potential for the OER process are β-GeSe (0.49 V)[6], C$_3$N$_5$ (0.69 V)[10], BeN$_2$ (0.83 V)[7], g-CN (0.93 V)[11], Pd$_3$P$_2$S$_8$ (1.07 V)[12], GaAs (1.39 V)[8], g-C$_3$N$_4$ (1.45 V)[91], CuCl (1.61 V)[13], Janus Pd$_4$S$_3$Se$_3$ (1.76 V)[27] and C$_3$S (2.03 V)[9].

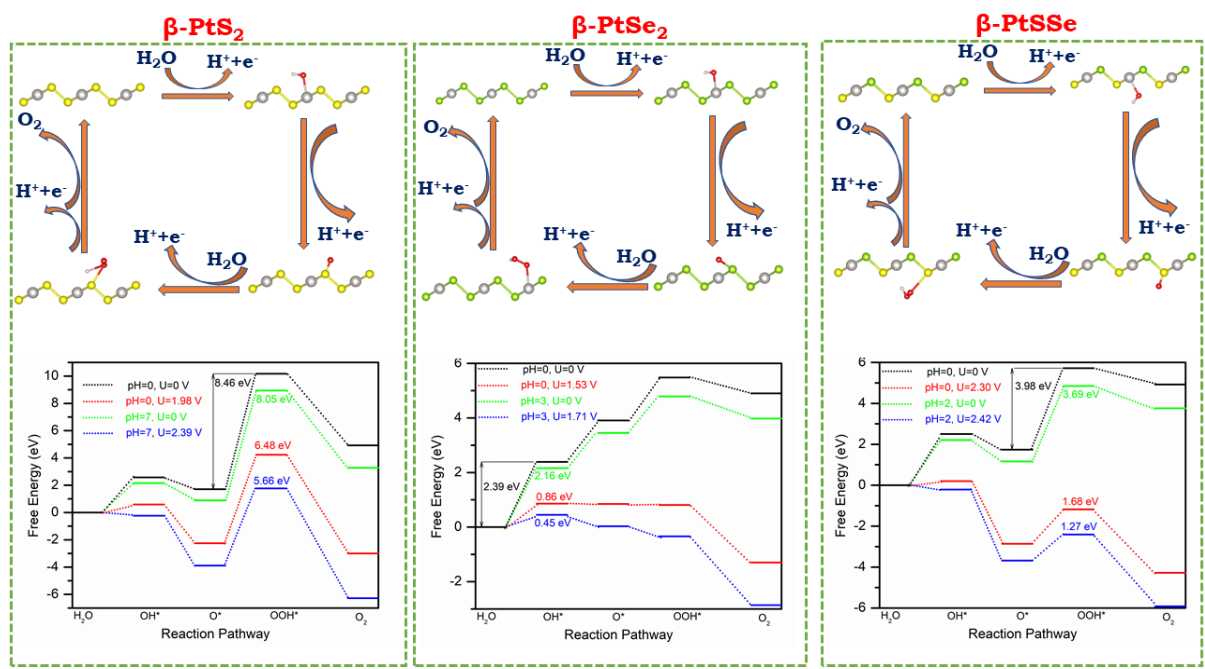

**Figure 5:** Photocatalytic pathways and free energy profiles of β-PtS$_2$, β-PtSe$_2$, and Janus β-PtSSe monolayers for OER at different pH photogenerated potential. Color code: Grey-Pt, Yellow-S, Green-Se, Light-H, Red-O.

## 3.6 Solar-to-Hydrogen Efficiency

Now, we calculate the solar-to-hydrogen efficiency, which is the product of light absorption and carrier utilization efficiency assuming 100% efficiency of the catalytic reaction[16]:

$$\eta_{STH} = \eta_{abs} \times \eta_{cu} = \frac{\int_{E_g}^{\infty} P(\hbar\omega)d(\hbar\omega)}{\int_{0}^{\infty} P(\hbar\omega)d(\hbar\omega)} \times \frac{\Delta G \int_{E}^{\infty}\frac{P(\hbar\omega)}{\hbar\omega}d(\hbar\omega)}{\int_{E_g}^{\infty} P(\hbar\omega)d(\hbar\omega)} \qquad (2)$$

Here $P(\hbar\omega)$ is AM1.5G solar energy flux at the photon energy $\hbar\omega$, and $E$ is the energy of photons. The bandgap and overpotentials of the hydrogen ($\chi(H_2)$), and oxygen ($\chi(O_2)$) are calculated using HSE06 hybrid functional (Table S4). The details are given in supplementary information. The calculated light absorption ($\eta_{abs}$), carrier utilization ($\eta_{cu}$), and STH ($\eta_{STH}$) efficiencies are listed in Table 4.

**Table 4:** The efficiency of light absorption ($\eta_{abs}$), carrier utilization efficiency ($\eta_{cu}$), solar-to-hydrogen efficiency ($\eta_{STH}$), and corrected solar-to-hydrogen efficiency ($\eta`_{STH}$) of β-PtS$_2$, β-PtSe$_2$, and Janus β-PtSSe monolayers.

|   | $\eta_{abs}$ (%) | $\eta_{cu}$ (%) | $\eta_{STH}$ (%) | $\eta`_{STH}$ (%) |
|---|---|---|---|---|
| β-PtS$_2$ | 5.0 | 41.0 | 2.1 | - |
| β-PtSe$_2$ | 33.5 | 48.1 | 16.1 | - |
| β-PtSSe | 31.3 | 59.5 | 18.6 | 17.7 |

The carrier utilization efficiency ($\eta_{cu}$) for all the monolayers are higher than 40 % due to appropriate levels of $\chi(H_2)$ and $\chi(O_2)$. Light absorption efficiency ($\eta_{abs}$) is highly dependent on the bandgap of photocatalysts materials. A relatively higher value of bandgap of β-PtS$_2$ (2.60 eV) as compared to β-PtSe$_2$ (1.77 eV) and β-PtSSe (2.09 eV) results in a low value of light absorption ($\eta_{abs}$= 5%) for β-PtS$_2$. It further leads to its low STH efficiency ($\eta_{STH}$ = 2.1%), which is lower than the critical value ($\eta_{STH}$ > 10%) for economic photocatalytic hydrogen production via photocatalysis[92]. However, the efficiency of light absorption of β-PtSe$_2$ and Janus β-PtSSe is more than 30% due to the appropriate band gap that leads to $\eta_{STH}$=16.1% for β-PtSe$_2$ and $\eta_{STH}$=18.6% for Janus β-PtSSe.

Also, the intrinsic electric field in the Janus structure eases the electron-hole separation in photocatalytic water splitting, which needs to be included in the total energy. The STH efficiency of 2D photocatalyst material with intrinsic electric field is corrected as[19]:

$$\eta'_{STH} = \eta_{STH} \ \frac{\int_{E}^{\infty} P(\hbar\omega)d(\hbar\omega)}{\int_{E_0}^{\infty} P(\hbar\omega)d(\hbar\omega) + \Delta\Phi \int_{E_g}^{\infty}\frac{P(\hbar\omega)}{\hbar\omega}d(\hbar\omega)} \qquad (3)$$

where ΔΦ is the electrostatic potential difference of Janus monolayer, which is 0.37 eV for β-PtSSe. The corrected $\eta_{STH}$ for Janus β-PtSSe is calculated to be 17.7%. Note that the $\eta_{STH}$ of β-PtSe$_2$ and Janus β-PtSSe is much higher than the recently reported 2D materials such as Ga$_2$S$_3$ (6.4%)[16], Ga$_2$SSe bilayer (7.42%)[15], pentagonal PdSe$_2$ (12.59 %)[3] and Janus WSSe (14.46%)[20].

## 4. CONCLUSIONS

In summary, we have investigated β-PtX$_2$ (X=S, Se) monolayers and their Janus structure β-PtSSe as photocatalysts for solar water splitting by employing the first-principles method. The results based on AIMD simulations and phonon dispersion reveal the thermal and kinetic stability of these monolayers. The cleavage and cohesive energy calculations suggest the possibility of experimental synthesis. The anisotropic high carrier mobility and good visible light harvesting ability of these monolayers make them a potential candidate for water splitting using sunlight. Remarkably, a small external potential of 0.67 V, 0.76 V, and 0.69 V is required to derive the HER process on the surface of β-PtS$_2$, β-PtSe$_2$, and Janus β-PtSSe at pH=0. Also, β-PtSe$_2$ and Janus β-PtSSe are OER active and require an external potential of 0.45 V and 1.27 V, respectively, to overcome the OER barrier. High light absorption and carrier utilization efficiency results in high solar-to-hydrogen energy conversion efficiency of β-PtSe$_2$ (~16 %) and Janus β-PtSSe (18%), which suggests their excellent photocatalytic activity for solar water splitting.

## ACKNOWLEDGEMENTS

PJ gratefully acknowledges the UGC for D. S. Kothari Post-Doctoral Fellowship. Superior, a high-performance computing cluster at Michigan Technological University, MI, Houghton, was used to obtain the results presented in this paper.